\documentclass{PoS}
\usepackage{epsfig}

\title{How accurate is the local-duality model for the pion elastic form factor?}

\ShortTitle{How accurate is the local-duality model for the pion elastic form factor?}

\author{\speaker{Irina Balakireva}
\\D.~V.~Skobeltsyn Institute of
Nuclear Physics, Moscow State University, 119991, Moscow, Russia
\\E-mail: \email{iraxff@mail.ru}}

\abstract{
We study the accuracy of the pion form factor, obtained with a local-duality version of dispersive sum rules. 
To probe this accuracy, we make use of a potential model, where the exact form factor may be calculated from 
the solution of the Schr\"odinger equation and confronted with the local-duality form factor. 
The deviation between these quantities is found to be below 20\% in the region of momentum transfers 
$Q\,>\,2\,-\,3\,\mbox{GeV}$, independently of the specific form of the confining potential. 
We argue that the local-duality model for elastic form factors in QCD has at least this level of accuracy.}

\FullConference{The XIXth International Workshop on High Energy Physics and Quantum Field Theory, QFTHEP2010\\
		September 8-15, 2010\\
		Golitsyno, Moscow, Russia}

\begin{document}
\section{Introduction}
\noindent
The pion elastic form factor at $Q\,=\,3\,-\,8\,\mbox{ GeV}$ is sensitive to the specific features of the onset of 
the perturbative regime and opens the possibility to study subtle details of the pion structure. 
The experimental study of the pion form factor at the lower part of this region will become available with 
an upgrade of JLab in the next few years. Therefore it is now the right time to review the theoretical understanding 
of this quantity. 
%The theoretical description of the pion form factor in this region directly from QCD is a complicated problem. 
%For example QCD on the lattice, has so far technical limitations which do not allow us to obtain the corresponding results.
Several versions of the method of QCD sum rules may and have been applied to this problem. 
However, no conclusive results have been obtained and we still have a strong discrepancy between the results from 
different versions of QCD sum rules \cite{braguta,bakulev}.
The goal of the presented work was to study the accuracy of the pion form factor obtained from so-called local-duality (LD) 
version of QCD sum rules \cite{radyushkin}. Let us briefly remind the basic ideas leading to the LD model.

\section{Sum rule}
\noindent
The basic objects for sum-rule calculations of pion properties are the correlation functions
% Let us consider the two- and three-point correlation functions.
\begin{eqnarray}
\label{sr_pion}
\Pi\,\left(p^2\right)\,&=&\,\int\left<\Omega\left|T\,j(x)\,j^{\dagger}(0)\right|\Omega\right>\,e^{ipx}\,dx,\nonumber\\
\Gamma\,\left(p^2_1,\,p^2_2,\,q^2\right)\,&=&\,\int\left<\Omega\left|T\,j(x_1)\,J(0)\,j^{\dagger}(x_2)\right|\Omega\right>
\,e^{ipx_1-ipx_2}\,dx_1\,dx_2. 
\end{eqnarray}
Here $\Omega$ is the physical vacuum; $j(x)$ is a short-hand notation for the interpolating axial current $j_{5\,\alpha}(x)$ of 
the positively charged pion, $\left<\Omega\left|j_{5\,\alpha}(0)\right|\pi\,(p)\right>\,=\,i\,p_{\alpha}\,f_{\pi}$;   
%
%\begin{equation}
%\label{Vacuum_Omega_Pi}
%\left<\Omega\left|j_{5\,\alpha}(0)\right|\pi\,(p)\right>\,=\,i\,p_{\alpha}\,f_{\pi}
%\end{equation}
$J(0)$ denotes the electromagnetic current $J_{\nu}(0)$. For brevity, we omit Lorentz indices. 
In QCD these correlators may be calculated by applying the operator product expansion (OPE). Instead of the Green functions in 
the Minkowski space (\ref{sr_pion}), it is convenient the evolution operators in the Euclidean space, 
which emerge after performing the Borel transform $p^2\to \tau$, the parameter $\tau$ being related to the Euclidean time. 
The Borel 
transform leads to several improvements: 
(i) suppresses the contributions of the excited states; (ii) improves the convergence of the perturbative expansion; 
(iii) provides the necessary smearing required by quark-hadron duality. The Borel image of the two-point correlator 
has the form
\begin{eqnarray}
\label{SR_P2}
\Pi_{\rm OPE}\,(\tau)\,=\,\int_0^{\infty}\rho_{\rm pert}(s)\,e^{-s\,\tau}\,ds\,+\,\Pi_{cond}\,(\tau),\qquad 
\rho_{\rm pert}(s)\,=\,\rho_0(s)\,+\,\alpha_s\,\rho_1(s)\,+\,O(\alpha^2_s). 
\end{eqnarray}
Here $\rho_i(s)$ are the spectral densities of the two-point diagrams of the perturbation theory, 
$\Pi_{cond}(\tau)$ describes nonperturbative 
power corrections. Making use of the hadron intermediate states, for the two-point correlator we obtain 
\begin{equation}
\label{SR_P3}
\Pi(\tau)\,=\,f^2_{\pi}\,e^{-m^2_{\pi}\,\tau}\,+\,excited\, states.
\end{equation}
The first term in this expression corresponds to the pion contribution.

The double Borel transform $p^2_{1,2}\to \frac{\tau}{2}$ of the three-point function has the form 
\begin{eqnarray}
\label{SR_G2}
\Gamma_{\rm OPE}\,(\tau,\,Q)\,=
\,\int_0^{\infty}\,\int_0^{\infty}\Delta_{\rm pert}(s_1,\,s_2,\,Q)\,e^{-\frac{s_1+s_2}2 \tau}\,ds_1\,ds_2\,+\,\Gamma_{cond}\,(\tau,\,Q),
\nonumber\\
\label{Delta_pert}
\Delta_{\rm pert}(s_1,\,s_2,\,Q)\,=\,\Delta_0(s_1,\,s_2,\,Q)\,+\,\alpha_s\,\Delta_1(s_1,\,s_2,\,Q)\,+\,O(\alpha^2_s),
\end{eqnarray}
$\Delta_{\rm pert}(s_1,\,s_2,\,Q)$ is the double spectral density of the three-point diagrams of the perturbation theory. 
$\Gamma_{\rm cond}(\tau,\,Q)$ describes power corrections. Making use of the hadron intermediate states gives 
\begin{eqnarray}
\label{SR_G3}
\Gamma(\tau,\,Q)\,=\,F_\pi(Q)\,f^2_{\pi}\,e^{-m^2_{\pi}\,\tau}\,+\,excited\,states.
\end{eqnarray}
The key assumption of the method of sum rules is the {\it duality assumption} which says that 
the contribution of the excited states is dual to the high-energy region of the perturbative diagrams. 
Using this assumption, 
the sum rules take the form in the chiral limit of the vanishing quark masses 
%\noindent
%(i)\,for the two-point correlator:
\begin{eqnarray}
\label{SR_P4}
f^2_{\pi}\,&=&\,\int_0^{\bar s_{\rm eff}(\tau)}\rho_{\rm pert}(s)\,e^{-s\,\tau}\,ds\,+\,
\frac{\left<\alpha_s\,G^2\right>}{12\,\pi}\,\tau\,+\,\frac{176\,\pi\,\alpha_s\,\left<\bar q\,q\right>^2}{81}\,\tau^2\,+\,\cdots,\\
\label{SR_G4}
F_\pi(Q)\,f^2_{\pi}\,&=&
\int_0^{s_{\rm eff}(Q,\,\tau)}
\int_0^{s_{\rm eff}(Q,\,\tau)}
\,\Delta_{\rm pert}(s_1,\,s_2,\,Q)\,
e^{-\frac{(s_1+s_2)}{2}\tau}ds_1ds_2
\nonumber
\\
&+&
\frac{\left<\frac{\alpha_s}{\pi}\,G^2\right>}{24}\,\tau\,+\,\frac{4\,\pi\,\alpha_s\,
\left<\bar q\,q\right>^2}{81}\,\tau^2\,\left(13\,+\,Q^2\,\tau\right)\,+\,\cdots
\end{eqnarray}
These relations are the standard relations for the extraction of the hadron decay constants and form factors 
in the method of QCD sum rules.

Let us focus on Eq.~(\ref{SR_G4}). We would like to study the form factor at large $Q$. The form factor of a bound state 
should decrease with $Q$; however, the power corrections of the r.h.s. are polynomials in $Q$ and thus rise with $Q$. 
So, Eq.~(\ref{SR_G4}) cannot be directly used at large $Q$. There are two ways for considering the region of large $Q$: 
The first way is the resummation of power corrections, after which the resummed power correction decrease with $Q$. 
This may be done in a model-dependent way by making use of nonlocal condensates \cite{bakulev}. 
The second way is just to set the Borel parameter $\tau=0$; then all power corrections vanish and the remaining perturbative 
part decreases with $Q$. This version of sum rules is called a local-duality (LD) sum rule \cite{radyushkin}. 
In the LD limit one finds 
\begin{eqnarray}
\label{SR_P5}
f^2_{\pi}\,=\,\int_0^{\bar s_{\rm eff}}\rho_{\rm pert}(s)\,ds\,=\,\frac{\bar s_{\rm eff}}{4\,\pi^2}\,
\left(1\,+\,\frac{\alpha_s}{\pi}\right)\,+\,O\left(\alpha_s^2\right), 
\\
\label{SR_G5}
F_\pi(Q)\,f^2_{\pi}\,=\,\int_0^{s_{\rm eff}(Q)}\,\int_0^{s_{\rm eff}(Q)}\Delta_{\rm pert}(s_1,\,s_2,\,Q)\,ds_1\,ds_2.
\end{eqnarray}
The double spectral densities $\rho_{\rm pert}(s)$ and $\Delta_{\rm pert}(s_1,\,s_2,\,Q)$ are given by the perturbation theory; 
$f_{\pi}$ is known from the experiments. So, if we fix $s_{\rm eff}(Q)$, the form factor may be calculated.
The spectral densities have the following properties: at $Q\,\to\,0$ the spectral densities of two- and three-point functions 
are related to each other by the Ward identity
\begin{equation}
\label{Ward_identity}
\lim\limits_{Q\to 0}\Delta_i\left(s_1,\,s_2,\,Q\right)\,=\,\rho_i(s_1)\,\delta\left(s_1\,-\,s_2\right). 
\end{equation}
At $Q\,\to\,\infty$ explicit calculations give:
\begin{eqnarray}
\label{large_momentum}
\lim\limits_{Q\to \infty}\Delta_0\left(s_1,\,s_2,\,Q\right)\,\sim\,\frac{1}{Q^4}, \qquad 
\lim\limits_{Q\to \infty}\Delta_1\left(s_1,\,s_2,\,Q\right)\,=\,\frac{8\,\pi}{Q^2}\,\rho_0(s_1)\,\rho_0(s_2).
\end{eqnarray}
For the pion form factor, two rigorous properties are known:  
The normalization condition related to the current conservation 
\begin{equation}
\label{normalization}
F_{\pi}(0)\,=\,1. 
\end{equation}
The asymptotic behavior at large $Q$ due to the QCD factorization theorem
\begin{equation}
\label{factorization}
F_{\pi}(Q)\,=\,\frac{8\,\pi\,f^2_{\pi}\,\alpha_s}{Q^2}\,+\,\cdots. 
\end{equation}
Obviously, if we set 
\begin{equation}
\label{seff_small_lagre}
s_{\rm eff}(Q\,\to\,0)\,=\,\frac{4\,\pi^2\,f^2_{\pi}}{1\,+\,\frac{\alpha_s}{\pi}},\qquad
s_{\rm eff}(Q\,\to\,\infty)\,=\,4\,\pi^2\,f^2_{\pi}, 
\end{equation}
then the form factor obtained from the LD sum rule (\ref{SR_G5}) satisfies both of these rigorous properties. 
The two values of the $s_{\rm eff}$ at large and at small $Q$ are not far from each other! 
So, it is easy to construct an interpolation function $s_{\rm eff}(Q)$ for all $Q$ with the limiting values (\ref{seff_small_lagre}).

Now, we can formulate the {\it LD model for hadron elastic form factors:}\footnote{
Notice, however, that the model is not expected to work at small nonzero $Q$ as the OPE is not applicable here.} 
\begin{itemize}
\item[(i)] It is based on a dispersive three-point sum rule at $\tau\,=\,0$ (i.e. infinitely large Borel mass parameter).
In this case all power corrections vanish and the details of the non-perturbative dynamics are hidden in a single 
quantity - the effective threshold $s_{\rm eff}(Q)$.
\item[(ii)] It makes use of a model for $s_{\rm eff}(Q)$ based on a smooth interpolation between its values at $Q\,\to \,0$ 
determined by the Ward identity and at $Q\,\to \,\infty$ determined by factorization. 
Since these values are not far from each other, one believes the details of this interpolation to be not essential.
For instance, a self-consistent expression may be used \cite{braguta}:
\begin{equation}
\label{seff_approx}
s_{\rm eff}(Q)\,=\,\frac{ 4\,\pi^2\,f^2_{\pi}}{1\,+\,\frac{\alpha_s(Q)}{\pi}}.
\end{equation}
\end{itemize}
Thus, the only non-perturbative input for the LD model is the pion decay constant $f_{\pi}$.

Obviously, the LD model is an approximate model which does not take into account the details of the confinement dynamics, 
and it is important to understand its accuracy. Now, where this accuracy may be tested? 

The only property of theory relevant for this model is factorization of hard form factors. Therefore, the model may be 
tested in quantum mechanics for the case of the potential containing both the Coulomb and the confining interactions.
The spectral representation for the form factor and the decay constant in the LD limit are similar to those in QCD; 
the corresponding spectral densities can be calculated from the two- and three-point diagrams of the non-relativistic 
field theory.
To probe the sensitivity of the LD model to the details of the confining potential, we shall consider two different 
confining potentials 
\begin{equation}
\label{potentials}
V(r)\,=\,-\frac{\alpha}{r}\,+\,V_{conf}(r),\qquad V_{conf}(r)\,=\,\sigma\,r\quad (1)\quad
V_{conf}(r)\,=\,\frac{m\,\omega^2\,r^2}{2}\quad (2)
\end{equation}
and make use of the parameters relevant for hadron physics. The parameters are chosen such that the Schr\"odinger equation for both confining 
potentials leads to the same value of decay constant (i.e. $\Psi(r=0)$): 
%\begin{equation}
%\label{parameters}
$m\,=\,0.35\,\mbox{ GeV},\quad \omega\,=\,0.5\,\mbox{ GeV},\quad \sigma\,=\,0.168\,\mbox{ GeV},\quad \alpha\,=\,0.3.$
%\end{equation}
The exact form factors as obtained from the solution of the Schr\"odinger equation are different in these two models; 
however the LD model for the form factor, which depends only on the value of $\alpha$ and $\Psi(r=0)$ for both models is the same. 
Comparing the exact form factors and the LD form factors allows us to probe the accuracy of the LD model. 
\section{Numerical results}
\begin{figure}[b!]
\begin{center}
\begin{tabular}{cc}
\includegraphics[width=7cm]{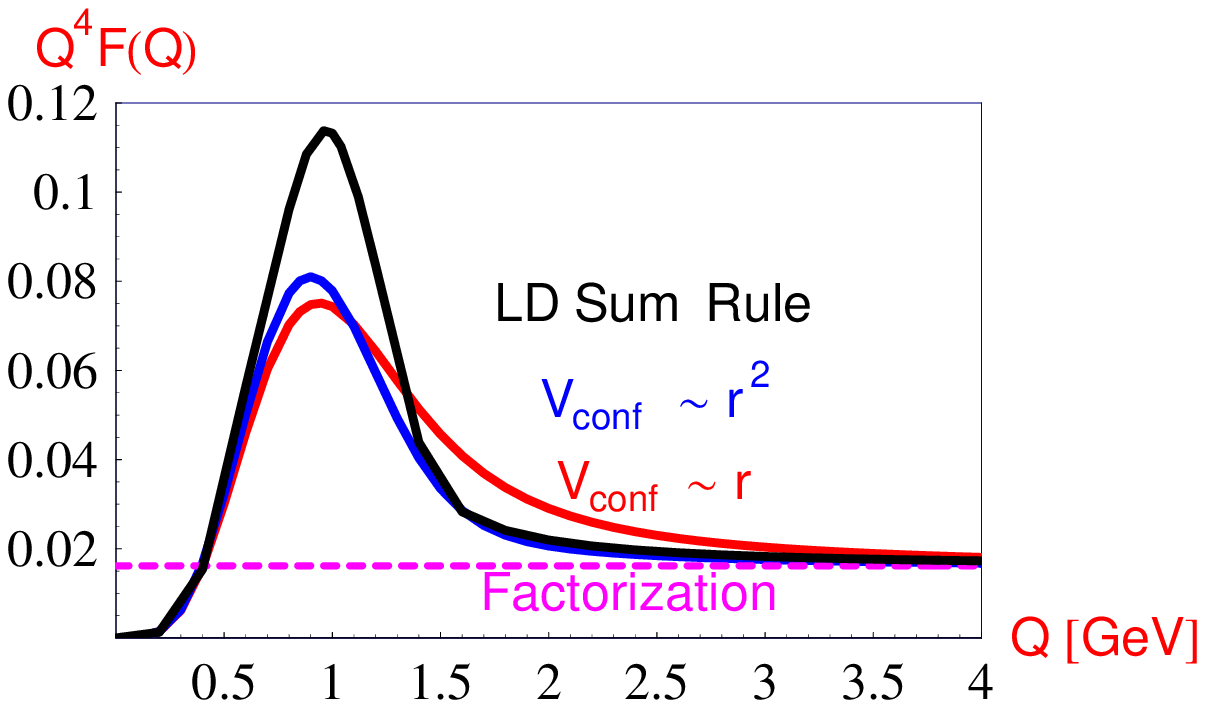}&
\includegraphics[width=7cm]{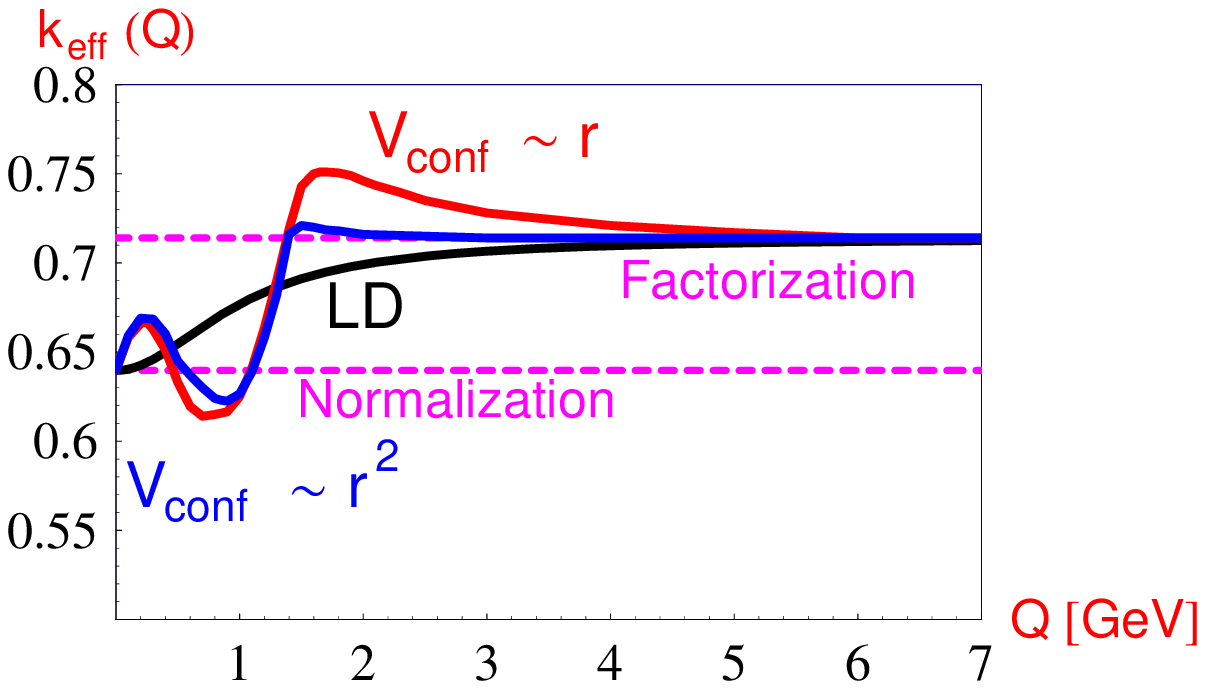}
\\
(a)& (b)
\end{tabular}
\caption{\label{Plot:1}
(a) The exact vs LD form factors. 
(b) The exact vs LD effective thresholds.
Red lines - linear confining potential; 
blue lines - harmonic oscillator confining potential; 
black lines - LD model. }
%\end{center}
%\end{figure}
%\
%\begin{figure}[b!]
%\begin{center}
\begin{tabular}{cc}
\includegraphics[width=7cm]{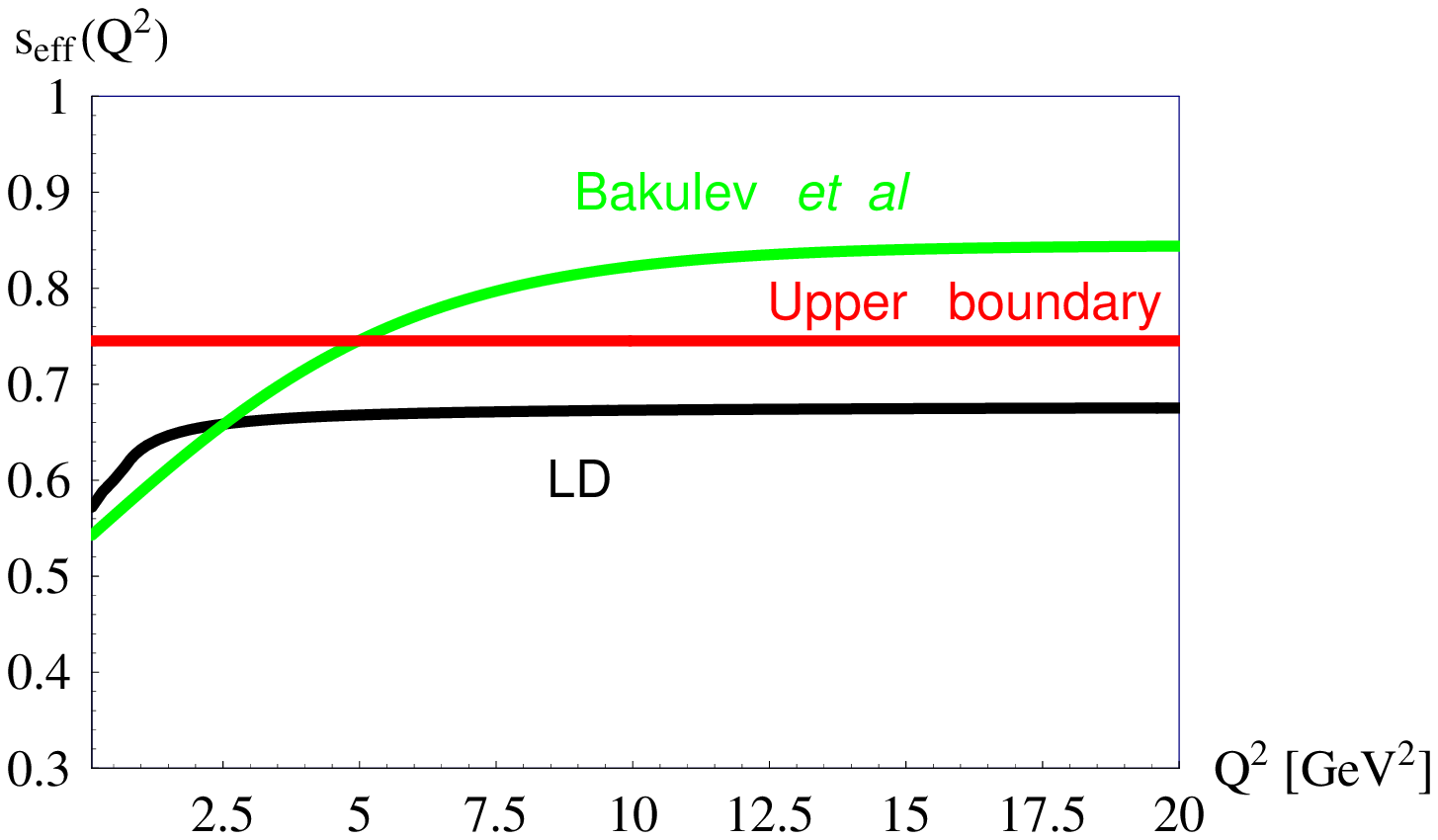}&
\includegraphics[width=7cm]{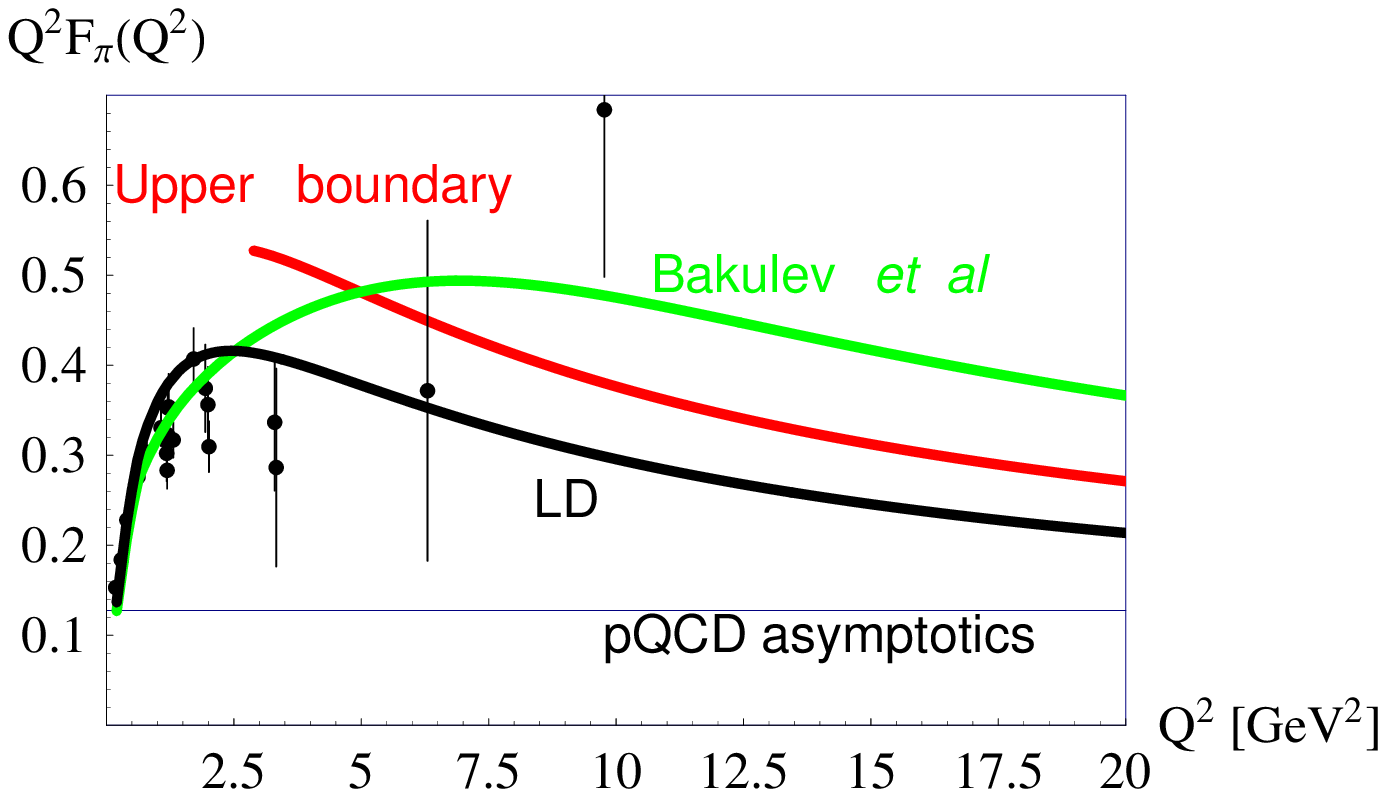}\\
(a)& (b)
\end{tabular}
\caption{\label{Plot:2}
(a) Models for the effective threshold. (b) The corresponding pion form factor. 
Black lines - the LD model. Red lines -  our upper boundary. 
Green lines - the results from the sum rule with non-local condensates \cite{bakulev}. Data from \cite{data_largeQ2}.}
\end{center}
\end{figure}

\vspace{-.2cm}
Fig.~\ref{Plot:1} presents our results for the potential model. 
The black lines give the results from the LD model for the effective threshold and the corresponding form factor. 
The blue and the red lines are the exact form factors obtained from the solution of the Schr\"odinger equation (Fig.~{Plot:1}a) 
and the corresponding exact thresholds which reproduce these form factors by the LD expression (\ref{SR_G5}). 
[The variable $k$ is related to the variable $s$, used above as $s\,=4(\,k^2\,+\,m^2)$].
Recall that the form factor in our potential model behaves as $F(Q^2)\sim 1/Q^4$ because we consider spinless quarks. 
The lesson to be learnt from the potential model is the following: {\it the exact threshold $k_{\rm eff}$ 
does not exceed the LD threshold by more than 5\%} As $Q$ increases, the accuracy of the LD approximation increases 
rather fast, too. 
This conclusion does not depend on the details of the confining interaction. 

Fig.~\ref{Plot:2} shows the analysis of the pion form factor in QCD. 
The black lines -- the results from the LD model; the red lines represent our upper boundary for the threshold and
the corresponding form factor. 
According to the experience from the potential model we set the upper boundary for the threshold $s_{\rm eff}$ 
by about 10\% higher than the asymptotic threshold. 
We believe this estimate to be quite reliable: the previous analysis of various correlators \cite{lms_sr1,lms_sr2} 
shows that the extraction procedures in quantum mechanics and in QCD are very similar both qualitatively and quantitatively. 
Notice also that the LD model for the pion form factor agrees well with the results from the dispersion approach \cite{anisovich}. 

The problem is that the effective threshold recalculated in \cite{bakulev} to reproduce the results from 
the sum rule with non-local condensates exceeds our upper boundary by more than 10\%. 
This might be a consequence of adopting in Ref.~\cite{bakulev} a procedure of fixing a $\tau$-independent 
effective threshold based on the maximal stability. According to the results of \cite{lms_sr1} this algorithm does not 
guarantee the extraction of reliable values. 

\section{Summary and Conclusions}
We studied the LD model for the elastic form factor which may be formulated in any theory where the form factor at large 
momentum transfers satisfies the factorization theorem (i.e., any theory containing both Coloumb and Confining interactions).

Our main conclusions are:
 
\noindent 
1. In the region $Q\,=\,1\,-\,2\,\mbox{ GeV}$, the exact effective threshold exhibits rapid variation with $Q$. 
Depending on the theory (relativistic or non-relativistic), the error of the LD form factor in this region may 
reach 30-40\% level. In general, for a relativistic theory a smaller error is expected.

\noindent 
2. At $Q\,>\,2\,-\,3\,\mbox{ GeV}$, the LD model provides a good description of the pion elastic form factor -- 
with the accuracy better than 20\%. Moreover, the accuracy increases rather fast with $Q$. 

We point out that our prediction for the pion form factor is considerably lower than the prediction of the approach
based on the sum rule with non-local condensates. This discrepancy needs clarification. 
Presumably \cite{melikhov}, its origin might be traced back to the procedure of fixing the $\tau$-independent 
effective threshold in the method of sum rules with non-local condensates based on merely the Borel stabilty criterion. 
We are going to redo the analysis making use of the recently formulated modifications of the sum-rule method based 
on the $\tau$-dependent effective threshold \cite{lms_sr2}.

\acknowledgments
I am grateful to Dmitri Melikhov for the supervision of this work and for valuable discussions. 

\vspace{-.2cm}
%\newpage

\end{document}